\documentclass[aps,prl,twocolumn,floatfix,a4paper,showpacs,superscriptaddress]{revtex4}
\usepackage{epsfig}
\usepackage{amsmath}
\usepackage{color}

\begin{document}
\title{Finding structural anomalies in graphs by means of quantum walks}
\author{Edgar Feldman}
\affiliation{Department of Mathematics, Graduate Center of the City University of New York, 365
Fifth Avenue, New York, NY 10016 USA}
\author{Mark Hillery}
\affiliation{Department of Physics, Hunter College of the City University of New York, 695 Park Avenue, New York, NY 10021 USA}
\author{Hai-Woong Lee}
\affiliation{Department of Physics, Korea Advanced Institute of Science and Technology, Daejeon
305-701,  Korea}
\author{Daniel Reitzner}
\affiliation{Research Center for Quantum Information, Slovak Academy of Sciences, D\'ubravsk\'a cesta 9, 845 11 Bratislava, Slovakia}
\author{Hongjun Zheng}
\affiliation{Department of Physics, Hunter College of the City University of New York, 695 Park Avenue, New York, NY 10021 USA}
\author{Vladim\'\i r Bu\v zek}
\affiliation{Research Center for Quantum Information, Slovak Academy of Sciences, D\'ubravsk\'a cesta 9, 845 11 Bratislava, Slovakia}

\begin{abstract}
We explore the possibility of using quantum walks on graphs to find structural anomalies, such as
extra edges or loops, on a graph.  We focus our attention on star graphs, whose edges are like
spokes coming out of a central hub.  If there are $N$ spokes, we show that a quantum walk can
find an extra edge connecting two of the spokes or a spoke with a loop on it in $O(\sqrt{N})$ 
steps.  We initially find that
if all of the spokes have loops except one, the walk will not find the spoke without a loop, but this
can be fixed if we choose the phase with which the particle is reflected from the vertex without the
loop.  Consequently, quantum walks can, under some circumstances, be used to find structural
anomalies in graphs.
\end{abstract}

\maketitle

A quantum walk is a quantum version of a random walk \cite{AhDaZa93}.  Both types of walks occur
on a graph, which is a set vertices connected by edges.  A particle making a quantum walk behaves
differently from one making a classical random walk, because the mathematical objects that govern
its motion are amplitudes rather than probabilities, and this means that interference effects will
play a role.  There are two basic types of quantum walks, one in which time progresses in discrete
steps \cite{AhAmKeVa01} and the other in which time is continuous \cite{FaGu98b}.  Here we shall 
be concerned with a particular version of the discrete-time walk known as the scattering quantum 
walk \cite{HiBeFe03}.  In this type of quantum walk the particle resides on the edges and scatters at 
the vertices at each time step.  Recently considerable experimental progress on implementing 
quantum walks has been made \cite{KnRoSi03} - \cite{ScCaPo09}, and recent review of the entire
subject can be found in \cite{kendon07}.

Quantum walks have been used to develop quantum algorithms, and this had proven to be a
fruitful approach \cite{ambainis07} - \cite{FaGoGu07}.  
They were first used to conduct searches on graphs \cite{ShKeWh03} - \cite{ReHiFeBu09}.  
In a quantum walk search the properties of one of the vertices differs from that of the others, often
by doing what the other vertices do but adding a sign flip, and this marks that vertex.  The
object of the search is to find the marked vertex.  Quantum walk searches have been explored on a
number of types of graphs, including, hypercubes, grids and complete graphs.  
Recently it has also been shown that 
quantum walks can find marked edges and a marked complete subgraph of a complete 
graph \cite{HiReBu09}.  Here we would like to explore a different question, whether quantum 
walks can find structural anomalies in graphs.  We will look
in some detail at the problem of finding an extra edge in a particular type of graph, and then present
results on finding other types of anomalous elements.

We will consider what we call a star graph.  This graph has a high degree of symmetry, which means
that analyzing walks on it becomes relatively simple, because the Hilbert space in which the walk
occurs is of relatively small dimension \cite{KrBr07,ReHiFeBu09}.
This graph has a central vertex, which we shall label 
$0$, and $N$ additional vertices, which we shall label $1$ through $N$.  The central vertex is 
connected to each of the other vertices by an edge, and, for now, the vertices $1,2,\ldots N$ are not
connected to each other by edges.  In order to construct a quantum walk on this graph we first need
a Hilbert space for the particle making the walk.  We specify this by means of an orthonormal basis
consisting of the states $\{ |0,j\rangle ,\  |j,0\rangle | j=1,2,\ldots N \}$.  The state $|0,j\rangle$
corresponds to the particle being on the edge between $0$ and $j$ going from $0$ to $j$, and the 
state $|j,0\rangle$ corresponds to the particle again being on the edge between $0$ and $j$, but
now going from $j$ to $0$. Next we need a unitary operator that advances the walk one time 
step.  That is provided by the collective action of unitaries at each vertex that tell how the 
particle scatters as it passes through that vertex.  If $U$ is the unitary that advances the walk one step,
it acts on a particle entering the vertex $0$ as
\begin{equation}
U|j,0\rangle = -r|0,j\rangle + t\sum_{k=1, k\neq j}^{N} |0,k\rangle ,
\end{equation}
where $r=(N-2)/N$ and $t=2/N$.  That is, the particle has an amplitude of $-r$ of being reflected
and an amplitude $t$ of being transmitted to one of the other edges.
We now need to choose what happens at the vertices $1$
through $N$.  If we make the choice $U|0,j\rangle = |j,0\rangle$ for $j>1$ and $U|0,1\rangle
=-|1,0\rangle$, we obtain an implementation of the Grover search algorithm.  Starting with
an equal superposition of all of the basis states, after $O(\sqrt{N})$ steps the particle will
be located on the edge connecting the vertices $0$ and $1$.

\begin{figure}
\includegraphics{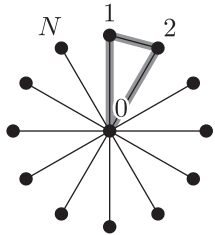}
\caption{A star graph with an extra edge between vertices 1 and 2.}
\label{star}
\end{figure}

Here we wish to do something different.  First, let's add an edge between vertices $1$ and $2$.  The
unitary operator will now act as  $U|0,j\rangle = |j,0\rangle$ for $j>2$, and
\begin{eqnarray}
U|0,1\rangle = |1,2\rangle & U|0,2\rangle = |2,1\rangle \nonumber \\
U|1,2\rangle = |2,0\rangle & U|2,1\rangle = |1,0\rangle .
\end{eqnarray}
Note that we have assumed that vertices $1$ and $2$ transmit the particle, and there
is no reflection.  One can put in an amplitude for reflection, but if it is not too large, this
does not change our results appreciably. 
The walk resulting from this choice of $U$ can be analyzed easily, because it stays within
a five-dimensional subspace of the entire Hilbert space.  Define the states
\begin{eqnarray}
|\psi_{1}\rangle & = & \frac{1}{\sqrt{2}} ( |0,1\rangle + 0,2\rangle ) \nonumber \\
|\psi_{2}\rangle & = & \frac{1}{\sqrt{2}} (|1,0\rangle + |2,0\rangle ) \nonumber \\
|\psi_{3}\rangle & = & \frac{1}{\sqrt{N-2}} \sum_{j=3}^{N} |0,j\rangle \nonumber \\
|\psi_{4}\rangle & = & \frac{1}{\sqrt{N-2}} \sum_{j=3}^{N} |j,0\rangle \nonumber \\
|\psi_{5}\rangle & = & \frac{1}{\sqrt{2}} (|1,2\rangle + |2,1\rangle ) .
\end{eqnarray}
These states span a five-dimensional space we shall call $S$.
The unitary transformation, $U$, that advances the walk one step acts on these states as follows:
\begin{eqnarray}
U|\psi_{1}\rangle & = & |\psi_{5}\rangle \nonumber \\
U|\psi_{2}\rangle & = & -(r-t) |\psi_{1}\rangle + 2\sqrt{rt} |\psi_{3}\rangle  \nonumber \\
U|\psi_{3}\rangle & = & |\psi_{4}\rangle  \nonumber  \\
U|\psi_{4}\rangle & = & (r-t) |\psi_{3}\rangle + 2\sqrt{rt} |\psi_{1}\rangle \nonumber \\
U|\psi_{5}\rangle & = & |\psi_{2}\rangle .
\end{eqnarray}
For our initial state we choose
\begin{eqnarray}
|\psi_{init}\rangle & = & \frac{1}{\sqrt{2N}} \sum_{j=1}^{N} ( |0,j\rangle - |j,0\rangle ) \nonumber \\
 & = & \frac{1}{\sqrt{N}} ( |\psi_{1}\rangle - |\psi_{2}\rangle ) \nonumber  \\
 & & + \sqrt{\frac{N-2}{2N}} ( |\psi_{3}\rangle -|\psi_{4}\rangle ) ,
 \end{eqnarray}
which is in $S$.  Since the initial state is in $S$, and $S$ is an invariant subspace of $U$, the
entire walk will remain in $S$, and this reduces the complexity of our problem considerably.  We
should mention that the minus sign in the first expression for initial state is essential; if it is replaced 
by a plus sign, the search will fail.  
 
In order to find the evolution of the quantum state for the walk, we find the eigenvalues and
eigenstates of $U$ restricted to $S$.  This gives us the spectral representation of $U$ and makes
finding $U^{n}$, the operator that will advance the walk $n$ steps, straightforward.
We then find that, to good approximation assuming that $N$ is large, 
 \begin{equation}
 U^{n}|\psi_{init}\rangle = \frac{(-1)^{n}}{\sqrt{3}}\left( \begin{array}{c} \sin (n\Delta ) \\  \sin (n\Delta ) \\
 \sqrt{3/2} \cos (n\Delta ) \\ -\sqrt{3/2} \cos (n\Delta ) \\ - \sin (n\Delta ) \end{array} \right) ,
 \end{equation}
 where the first entry is the coefficient of $|\psi_{1}\rangle$, the second is the coefficient of 
 $|\psi_{2}\rangle$, etc. , and $\Delta = (2t/3)^{1/2}$.  This is the state of the walk after $n$ steps.
From this equation, and the definitions of $|\psi_{1}\rangle$ through $|\psi_{5}\rangle$,
we see that when $n\Delta = \pi /2$, the particle is located on one of the
 edges leading to the extra edge or on the extra edge itself.  This will happen when $n = O(\sqrt{N})$.

We now need to discuss how to interpret this result.  It is reasonable to assume that if we are given
a graph with an extra edge in an unknown location, we only have access to the edges connecting
the central vertex to the outer ones, and not to the extra edge itself (if we had access to the extra 
edge, then we would have to know where it is).  That is, in making a measurement,
we can only determine which of the edges connecting central vertex to to the outer ones the
particle is on.  If it is on the extra  edge, we will not detect it.  So, after $n$ steps, where
 $n\Delta = \pi /2$, we measure the edges to which we have access to find out where the particle is.
 With probability $2/3$ it will be on an edge connected to the extra edge, and with probability $1/3$
 it will be on the extra edge itself, in which case we won't detect it.
 
 In comparing this procedure to a classical search for the extra edge, we shall assume that classically
 the graph is specified by an adjacency list, which is an efficient specification for sparse graphs.
 For each vertex of the graph, one lists the vertices that are connected to it by an edge.  In our
 case, the central vertex is connected to all of the other vertices, the vertices not connected to the
 extra edge are connected only to the central vertex, and two of the outer vertices are connected to
 the central vertex and to each other.  Searching this list classically would require $O(N)$ steps
 to find the extra edge, while the quantum procedure will succeed in $O(\sqrt{N})$.
 
 One can ask whether other structural anomalies can be detected by means of a quantum walk.
 One possibility is to take the basic star graph and add a loop to one of the outer vertices, say
 vertex $1$.  If we call the state corresponding to the particle being on the loop $|l_{1}\rangle$,
 the unitary time-step operator would act on the states going into the outer vertices as
 $U|0,1\rangle=|l_{1}\rangle$, $U|l_{1}\rangle = |1,0\rangle$, and $U|0,j\rangle = |j,0\rangle$
 for $j>1$.
 The action of $U$ for states going into the central vertex is as before.  The details of the 
 calculations for this walk will be presented elsewhere, but the result is similar to what we
 found in the case of the extra edge.  Starting with the same initial state as before, if 
 $n\sqrt{t/3}=\pi /2$, then with probability $2/3$ the particle will be on the edge connecting
 vertex $1$ to the central vertex, and with probability $1/3$ it will be on the loop itself.  Therefore,
 a quantum walk can be used to find the vertex with the loop attached in $O(\sqrt{N})$ steps.
 
A second possibility is simply to extend one of the edges.  One adds an extra edge and and extra
vertex, which we shall call $A$.  One end of this edge is attached to vertex $1$, and the other to
vertex $A$.  Vertex $A$ is connected only to vertex $1$.  The unitary time-step operator now
acts on the states going into the outer vertices as $U|0,1\rangle = |1,A\rangle$, $U|1,A\rangle
=|A,1\rangle$, $U|A,1\rangle = |1,0\rangle$, and $U|0,j\rangle = |j,0\rangle$ for $j>1$.  For an
initial state one chooses an arbitrary superposition of all of the states going out, 
$|\psi_{out}\rangle = (1/\sqrt{N})\sum_{j=1}^{N}|0,j\rangle$ and all of the states going in, 
$|\psi_{in}\rangle = (1/\sqrt{N}) \sum_{j=1}^{N}|j,0\rangle$.  For any initial state of this type, 
if one then runs a quantum walk on this graph, the particle does not become localized on the 
 extra edge or on the edge leading to it in $O(\sqrt{N})$ steps.  So, in this case, the quantum 
 walk fails to find the structural anomaly with a quantum speedup.  
 
Finally, let us see whether a quantum walk can find a missing element.  Suppose we add loops
to all of the outer vertices of our star graph except for one, say, as usual, vertex 1.  We shall 
designate the loop state connected to vertex $j$ by $|l_{j}\rangle$.  The action of the unitary
time-step operator on the outer vertices is now $U|0,1\rangle = |1,0\rangle$, and for $j>1$,
$U|0,j\rangle = |l_{j}\rangle$, and $U|l_{j}\rangle = |j,0\rangle$.  As in the previous case,  starting 
with a state that is an arbitrary superposition of the ingoing and outgoing states, we find that 
for no initial state of this type does the
particle become localized on the edge without the loop in $O(\sqrt{N})$ steps, so that 
the quantum walk again fails to find the anomaly with a quantum speedup.
 
These failures can be turned into successes, however, if we make a small modification to the walks.
In the case of the extra edge, suppose that instead of $U|1,A\rangle = |A,1\rangle$, we have 
$U|1,A\rangle = -|A,1\rangle$, with the action of $U$ on all of the other states being the same as
before.  Then with an initial state that is an arbitrary superposition of $|\psi_{out}\rangle$ and 
$|\psi_{in}\rangle$, the particle will be localized on the extra edge and the edge leading to it in 
$O(\sqrt{N})$ steps.  This situation is very much reminiscent of the standard Grover search.  

The case of the missing loop is more interesting.  
Suppose that instead of $U|0,1\rangle = |1,0\rangle$, we have that $U|0,1\rangle
= e^{i\phi} |1,0\rangle$.  We find that the particle will be localized on the edge with the missing loop
in $O(\sqrt{N})$ steps, if $\phi = \pi , \pm (\pi /3)$, and if the initial state is properly chosen.  In order
to explain this last point, let us rephrase the problem slightly.  We will add a dummy loop to the
vertex $1$, $|l_{1}\rangle$, where $U|l_{1}\rangle = |l_{1}\rangle$, that does not participate in the
dynamics  (we still have $U|0,1\rangle = e^{i\phi} |1,0\rangle$).  
So our search problem becomes finding the vertex with the dummy loop.  Define the state
$|\psi_{loop}\rangle = (1/\sqrt{N})\sum_{j=1}^{N} |l_{j}\rangle$.  Now, for the case $\phi = \pi$ we will
have a successful quantum walk search, i.e. the particle will become localized on the edge connected
to the dummy loop in $O(\sqrt{N})$ steps, if we start in the state
\begin{equation}
|\psi_{init}\rangle = \frac{1}{\sqrt{3}}( |\psi_{out}\rangle + |\psi_{in}\rangle + |\psi_{loop}\rangle )
\end{equation}
and in the case $\phi = \pi /3$ the proper initial state is
\begin{eqnarray}
|\psi_{init}\rangle & = & \frac{1}{1-e^{2\pi i/3} } (e^{-2\pi i/3}|\psi_{out}\rangle + |\psi_{in}\rangle
\nonumber \\
 & & + e^{2\pi i/3} |\psi_{loop}\rangle ) .
\end{eqnarray} 
Therefore, by adjusting the phase on the edge with the dummy loop and choosing the proper initial
state, the quantum walk can find the location of the dummy loop in $O(\sqrt{N})$ steps.

We have found that a successful quantum walk search on a modified star graph is associated
with a degenerate eigenvalue of the unperturbed problem.  The unperturbed evolution operator,
$U_{0}$, is obtained by setting $r=1$ and $t=0$ in matrix for $U$ (this is the $N\rightarrow\infty$
limit of $U$).  The perturbation, 
$\Delta U=U-U_{0}$ is small if $N$ is large.  If an eigenvalue of $U_{0}$ is simple,  we have found
that adding the perturbation adds a correction to it of order $1/N$, while if it is degenerate, adding
$\Delta U$ removes the degeneracy, and one obtains corrections of order $1/\sqrt{N}$.  Note
that $U^{n}$ can be expressed as
\begin{equation}
U^{n} = \sum_{j} \lambda_{j}^{n} P_{j} ,
\end{equation}
where $\lambda_{j}$ is an eigenvalue of $U$ and $P_{j}$ is the projection onto the corresponding
eigenvector.  Because $U$ is unitary, its eigenvalues have a magnitude of one.  
Now suppose that $\lambda_{j} = \exp [i(\theta_{j} + \Delta\theta_{j})]$, where
$ \exp (i\theta_{j})$ is the corresponding eigenvalue of $U_{0}$ and $\exp (i\Delta\theta_{j})$
is the correction due to $\Delta U$.  For the quantum search to succeed in $O(\sqrt{N})$ steps,
i.e. $n= O(\sqrt{N})$, the effect of $\Delta U$ on the state, which is given by 
$\exp (in\Delta\theta_{j})$, must be significant.  This will be true if $\Delta \theta_{j} = O(1/\sqrt{N})$,
but not if $\Delta \theta_{j} = O(1/N)$.

In conclusion, we have shown that quantum walks can find structural anomalies in graphs, and not
just marked elements.  Here, only a few examples have been studied, so the question of what
kinds of anomalies on what kinds of graphs can be efficiently found by means of a quantum
walk is largely open.  For example, it has been found that if one removes an edge from a 
complete graph, merely adjusting the the reflection and transmission amplitudes of the affected
vertices to maintain unitarity, a quantum walk will not efficiently find the missing edge \cite{Hi10}.
This leads one to ask whether there are structural changes one can make in a complete graph that
will be efficiently found in a quantum walk search.  Questions such as this remain for the future.  
However, what we have found here suggests that the types of objects that can be found by a 
quantum walk search go beyond just marked vertices.

\section*{Acknowledgments}
Our work has been supported by projects QAP 2004-IST-FETPI-15848, HIP FP7-ICT-2007-C-221889, APVV QIAM, CE SAV QUTE and by the National Science Foundation under grant number PHY-0903660.

\end{document}